\documentclass[12pt]{article}
\usepackage{amssymb,amsmath}
\usepackage{cite}
\usepackage{verbatim}
\usepackage{graphicx}

\usepackage{bm}

\let\OLDthebibliography\thebibliography
\renewcommand\thebibliography[1]{
  \OLDthebibliography{#1}
  \setlength{\parskip}{0pt}
  \setlength{\itemsep}{0pt plus 0.3ex}
}

\newcommand{\dd}{\mbox{\rm d}}

\newcommand{\gam}{\gamma}
\newcommand{\Gam}{\Gamma}

\newcommand{\tl}{\tilde}

\newcommand{\sx}{\mathring{X}}
\newcommand{\sX}{\mathring{X}}
\newcommand{\sq}{\mathring{q}}

\newcommand{\mr}{\mathring}

\newcommand{\DD}{\mbox{\rm D}}

\newcommand{\p}{\partial}

\newcommand{\be}{\begin{equation}}
\newcommand{\bear}{\begin{eqnarray}}
\newcommand{\ear}{\end{eqnarray}}
\newcommand{\ee}{\end{equation}}
\newcommand{\lbl}{\label}
\newcommand{\bi}{\bibitem}
\newcommand{\ci}{\cite}

\newcommand{\vs}{\vspace}
\newcommand{\hs}{\hspace}

\newcommand{\bp}{{\bm p}}

\newcommand{\bx}{{\bm x}}
\newcommand{\bX}{{\bm X}}
\newcommand{\bN}{{\bm N}}

\begin{document}

\

\baselineskip .7cm 

\vs{8mm}

\begin{center}

{\LARGE On the Coupling of Relativistic Particle to Gravity and Wheeler-DeWitt Quantization}

\vs{3mm}

Matej Pav\v si\v c

Jo\v zef Stefan Institute, Jamova 39,
1000 Ljubljana, Slovenia

e-mail: matej.pavsic@ijs.si

\vs{6mm}

{\bf Abstract}
\end{center}

\baselineskip .5cm 

{\footnotesize
A system consisting of a point particle coupled to gravity is investigated. The set of constraints
is derived. It was found that a suitable superposition of those constraints is the generator
of the infinitesimal transformations of the time coordinate $t\equiv x^0$ and serves as the Hamiltonian
which gives the correct equations of motion. Besides that, the system satisfies the mass shell
constraint, $p^\mu p_\mu - m^2 = 0$, which is the generator of the worldsheet reparametrizations,
where the momenta $p_\mu$, $\mu=0,1,2,3$, generate infinitesimal changes of the particle's position
$X^\mu$ in spacetime. Consequently, the Hamiltonian contains $p_0$, which upon quantization becomes
the operator $- i \p/\p T$, occurring on the r.h.s. of the Wheeler-DeWitt euqtion. Here the role
of time has the particle coordinate $X^0 \equiv T$, which is a distinct concept than the spacetime
coordinate $x^0 \equiv t$. It is also shown how the ordering ambiguities can be avoided if a
quadratic form of the momenta is cast into the form that instead of the metric contains the basis
vectors.
}

\baselineskip .6cm

\section{Introduction}

When quantizing gravity, one faces a tough problem, because time disappears from the equations.
If gravity is coupled to matter, then the changes of matter configurations are supposed to have
the role of time in quantum gravity (see, e.g, a review by Anderson\ci{Anderson}). Typically,
matter is described by scalar, spinor or electromagnetic fields\ci{Kiefer}. A different approach was
explored by Rovelli\ci{Rovelli1,Rovelli2} who considered as a model a single particle coupled to general
relativity. In addition to particle's coordinates $X^\mu (\tau)$, he also considered a clock
variable, attached to the particle. In Ref.\ci{PavsicKleinWheeler} a model without the clock variable
was investigated and it was found that the particle's coordinates $X^0$ (as well as $X^i$, $i=1,2,3$)
survive quantization and has the role of time in quantum gravity in the presence of the particle.
The model was also extended to a system of particles\ci{PavsicKleinWheeler}
and further elaborated\ci{PavsicKleinWheeler1}.
Recently that model was reconsidered by Struyve\ci{Struyve1,Struyve2}. He put the action into such a form that
the matter and gravity part had the same time parameter $\tau$. This required to insert an extra
${\dot X}^0$ into the gravity part of the action, and thus change the canonical momenta and
the constraint. Instead of the usual mass shell constraint $p^\mu p_\mu - m^2 = 0$, he obtained a new,
more complicated constraint that contained the Ricci scalar $R$. With this new constraint, it turned
out that upon quantization the time parameter $\tau$ disappeared from the equations. But Struyve
also observed that by a suitable canonical transformation at the classical level and a unitary 
transformation at the quantum level one can arrive at the equations obtained
in Ref.\ci{PavsicKleinWheeler,PavsicKleinWheeler1}.

In the present work we intend to clarify this important subject. Firstly, we observe that both the particle and
gravity part of the action can be cast into the form in which they both have the same evolution
parameter, namely, $t\equiv x^0$, while retaining the particle worldline parameter $\tau$ and the
mass shell constraint $p^\mu p_\mu - m^2 =0$. Rewriting the total action by employing the
ADM (1+3) split and varying it with respect to the lapse and shift (considered as Lagrange multipliers),
one obtains the constraints. As a Hamiltonian we take a superposition of those constraints and
find that it leads to the correct equations of motion (the geodesic equation and the
Einstein equations) by emplying the ordinary Poisson brackets. By this we varify the correctness of the Hamiltonian so constructed.

To further explore the meaning of the quantities such as the particle's momenta $p_\mu$ and the Hamiltonian,
we perform the total variation of the action that includes a change $\delta x^\mu$ and $\delta \tau$ of the boundary. So we obtain the generator $H$ of the transformations $\delta t$, the generators $p_\mu$ of
the transformations $\delta X^\mu$ (which are changes of particle's position in spacetime), and
the generator of  the transformation $\delta \tau$ (which is proportional to the mass shell constraint).
Such fundamental analysis, at each step covariant, convinces us that all the
momenta $p_\mu$, $ \mu = 0,1,2,3$, take place within the formalism. At the classical level, the presence of
the particle enables the identification (definition) of spacetime points. At the quantum level those
particle variables $X^\mu$, including $X^0$, remain in the equations; the particle's coordinate $X^0$
has the role of time. Because of the presence of the mass shell constraint, the Hamilton obtained by
Rovelli, namely\footnote{
	We omit here the clock variable that is included in the Rovelli's equation.}
\be
  H = \int \dd^3 \bx \,N^\mu {\cal H}_\mu^{\rm ADM} - N^i p_i - N \sqrt{m^2 + \bp^2} ,
\lbl{1.1}
\ee
can be written as $H= \int \dd^3 \bx \, N^\mu {\cal H}_\mu^{\rm ADM} - p_0$. Upon quantization,
$p_0 \rightarrow {\hat p}_0 = - i \frac{\p}{\p X^0}$. We see that the presence of the particle
``saves'' the concept of spacetime, so that time, namely, $X^0$, is present both in the classical and
quantum equations.
Otherwise it would be difficult to retain in the quantum theory the concept of local Lorentz
transformations of Eq.\,(\ref{1.1}), and understand how different local inertial observers
compare the observed values of $p_i$ without bringing $p_0$ into the description.

In Sec.\,2 we first point out that the Einstein equations imply the relation
$\frac{1}{8 \pi G} \int \dd^3 \bx \sqrt{-g} {G_0}^0 = - p_0$. Then we show that the analogous equation
comes out in the ADM formalism. In Sec.\,3 we discuss quantization of that model. At the end we also
touch the problem of the ordering ambiguities and point out how they could be avoided.

\section{Gravity coupled to particle}

The action for particle coupled to the gravitational field is\footnote{
	Such action makes sense if $X^\mu (\tau)$ are not meant to be the coordinates of an exactly
	point particle, but coordinates of the center of mass of an extended object. Here
	we thus describe not a point particle, but an extended particle (object) coupled to
	gravity, and include into the description only a restriced set of the object's variables,
	namely its center of mass coordinates.}
\be
  I[X^\mu,g_{\mu \nu}] = m \int \dd \tau \left (g_{\mu \nu} {\dot X}^\mu {\dot X}^\nu \right )^{1/2} + \frac{1}{16 \pi G} \int \dd^4 x \, \sqrt{-g}\, R  .
\lbl{2.1}
\ee
The variation of this action with respct to the metric $g_{\mu \nu}$ gives
\be
  - \frac{1}{8 \pi G}\, G^{\mu \nu}
     = \int \dd \tau \,\delta^4 (x-X(\tau))\frac{m {\dot X}^\mu {\dot X}^\nu}  {(g_{\alpha \beta} {\dot X}^\alpha {\dot X}^\beta)^{1/2} \sqrt{-g}} = T^{\mu \nu} ,
\lbl{2.2}
\ee
which are the Einstein equations in the presence of the stress-energt tensor $T^{\mu \nu}$ of the
particle.

From Eq.\,(\ref{2.2}) we obtain
\be
    - \frac{1}{8 \pi G}\, \int G^{\mu \nu} \sqrt{-g} \,\dd \Sigma_\nu
     = \int T^{\mu \nu} \sqrt{-g} \,\dd \Sigma_\nu = p^\mu ,
\lbl{2.3}
\ee
where $p^\mu$ is the particle's momentum. Writing the hypersurface element as
$\dd \Sigma_\nu = n_\nu \dd \Sigma$ and taking coordinates such that $n_\nu = (1,0,0,0)$ and
$\dd \Sigma = \dd^3 \bx$, we have
\be
- \frac{1}{8 \pi G}\, \int G^{\mu 0} \sqrt{-g}\, \dd^3 \bx
= \int T^{\mu 0} \sqrt{-g}\, \dd^3 \bx = \frac{m {\dot X}^\mu}{\sqrt{{\dot X}^\alpha {\dot X}_\alpha}} .
\lbl{2.4}
\ee
Here we have used $\int \dd \tau f(\tau) \delta (x^0 - X^0 (\tau))$ $= \frac{f(\tau)}{{\dot X}^0}|_{\tau_c}$, where $\tau_c$ is the solution of the equation $x^0 = X^0 (\tau)$.

Because of the Bianchi identity, ${G^{\mu \nu}}_{;\nu}=0$ (implying ${T^{\mu \nu}}_{;\nu}=0$),
not all equations (\ref{2.2}) are independent. The equations $\frac{1}{8 \pi G} G^{0\nu} + T^{0 \nu}=0$
are constraints on initial data, and so are the equations $\frac{1}{8 \pi G} G_{0\nu} + T_{0 \nu}=0$.
We thus have four constraints
\be
  \phi_\nu = \frac{1}{8 \pi G} G_{0\nu} + T_{0 \nu}=0 .
\lbl{2.5}
\ee
Similarly, not all components of the metric $g_{\mu \nu}$ are independent. The components $g_{0 \nu}$
can be chosen to be an artifact of a choice of coordinates and to have the role of Lagrange multipliers.
The same holds for $g^{0 \nu}$. The variation of the action (\ref{2.1}) with respoct to
$g^{0 \nu}$ (having the role of Lagrange multipliers) gives the constraints (\ref{2.5}).

A linear superposition of the constraints (\ref{2.5}) is the Hamiltonian:
\be
   H = \int \alpha^\nu \phi_\nu \, \sqrt{-g} \,\dd^3 \bx
    = \int \left ( \frac{1}{8 \pi G} G_{0\nu} + T_{0 \nu}\right ) g^{0 \nu} \sqrt{-g}\, \dd^3 \bx = 0,
\lbl{2.6}
\ee
where $\alpha^\nu = g^{0 \nu}$ are arbitrary functions of $x^\mu$. We thus have
\be
   \int \frac{1}{8 \pi G} G_{0\nu} g^{0 \nu} \sqrt{-g} \,\dd^3 \bx = - p_0 ,
\lbl{2.7}
\ee
where
\be
   p_0 = \int {T_0}^\nu \sqrt{-g} \,\dd \Sigma_\nu = \int {T_0}^0 \sqrt{-g} \,\dd^3 \bx
   = \frac{\p L}{\p {\dot X}^0}
   = - \frac{m \, g_{0 \mu} {\dot X}^\mu }{\sqrt{g_{\alpha \beta} {\dot X}^\alpha \, {\dot X}^\beta}}.
\lbl{2.7a}
\ee

The phase space form of the action (\ref{2.1}) is
$$
  I[X^\mu,p_\mu,\pi^{ij},q_{ij},\alpha,N,N^i] = \int \dd \tau \left [ p_\mu {\dot X}^\mu
  - \frac{\alpha}{2} (g^{\mu \nu} p_\mu p_\nu - m^2) \right ] \delta^4 (x-X(\tau)) \,\dd^4 x $$
\be
  \hs {2cm} + \int \dd^4 x \left (\pi^{ij} \sq_{ij}
   - N {\cal H}^{\rm ADM} + N^i {\cal H}_i^{\rm	ADM} \right ),
\lbl{2.8}
\ee
where $\pi^{ij}$, $q_{ij}$,$i,j = 1,2,3$, are the ADM phase space variables\ci{ADM1,ADM2},
and ${\cal H}^{\rm ADM}$,
${\cal H}_i^{\rm ADM}$ the ADM expressions for the gravitation part of the constraints. Here
${\dot X}^\mu \equiv \dd X^\mu/\dd \tau$ and $\sq_{ij} \equiv \dd q_{ij}/\dd t$. Later we will also
have $\sx^i \equiv \dd X^i/\dd t$, .
 
Because of the inserted $\delta$-function,tha matter and the gravitational part of the action have
both the same time parameter $x^0 \equiv t$.

Performing the integration overs $\tau$, we obtain
\be
  I= \int \dd t \, \dd^3 \bx \left [ \frac{\delta^3 (\bx-\bX (t))}{{\dot X}^0}
  \left ( p_\mu {\dot X}^\mu - \frac{\alpha}{2}  
  \left ( g^{\mu \nu} p_\mu p_\nu - m^2 \right ) \right ) \bigg|_{\tau_c} 
  + \pi^{ij} \sq_{ij} - N {\cal H}^{\rm ADM} + N^i {\cal H}_i^{\rm ADM} \right ] ,
\lbl{2.9}
\ee
where $\tau_c$ is the solution of the equation $x^0 - X^0 (\tau) = 0$. Expressing the metric
according to the ADM split\ci{ADM1,ADM2},
\be
g_{\mu \nu} = \begin{pmatrix}  N^2-N^i N_i , & -N_i& \\
	-N_j ,& - q_{ij} \\
\end{pmatrix}~,~~~~
g^{\mu \nu} = \begin{pmatrix}  {1}/{N^2} , & -{N^i}/{N^2}& \\
	-{N^j}/{N^2} ,& {N^i N^j}/{N^2}-q^{ij}\\
\end{pmatrix}
\lbl{2.9a}
\ee 
where $N=\sqrt{1/g^{00}}$ and $N_i = - g_{0 i}$, $i=1,2,3$, are the laps and
shift functions, we have
$$
I= \int \dd t \, \dd^3 \bx \left [ \delta^3 (\bx-\bX (t))
\left ( p_0 + p_i \sX^i - \frac{\alpha}{2{\dot X}^0}  
\left ( \frac{1}{N^2}(p_0 - N^i p_i)^2-q^{ij}p_i p_j - m^2 \right ) \right )\bigg|_{\tau_c} \right .$$
\be
   \hs{2cm}+ \left .\pi^{ij} \sq_{ij} - N {\cal H}^{\rm ADM} +   N^i {\cal H}_i^{\rm ADM} \right ] ,
\lbl{2.10}
\ee
Here we identify  $\alpha/{\dot X}^0$ with a new Lagrange multiplier according to
$\alpha/{\dot X}^0 = \lambda$, because ${\dot X}^0$ is arbitrary and has no longer a formal
role of a velocity as it had in the original action (\ref{2.8}).

The variation of this action with respect to the 3-metric $q_{ij}$ gives the $(ij)$-components
of the Einstein equation in the ADM split. The variation with respect to other variables gives:
\bear
  \delta p_0 &:&1 = \frac{\alpha}{{\dot X}^0}\frac{1}{N^2}(p_0 - N^i p_i) = 
  \frac{\alpha}{{\dot X}^0} p^0 ~~ \Rightarrow~~ p^0 = \frac{\alpha}{{\dot X}^0} ,
\lbl{2.11}\\
  \delta p_i &:&\sX^i = \frac{\alpha}{{\dot X}^0} \frac{N^i}{N^2}(p_0 - N^j p_j) - q^{ij}p_j
  = p^i \frac{\alpha}{{\dot X}^0}~~ 
   \Rightarrow~~ p^i = \frac{\sX^i {\dot X}^0}{\alpha} = \sX^i p^0 ,\lbl{2.12}\\
 \delta \alpha &:&   \frac{1}{N^2}(p_0 - N^i p_i)^2-q^{ij}p_i p_j - m^2 = 0, \lbl{2.15}
\ear
\be
\Rightarrow~~ p_0 - N^i p_i = \pm N \sqrt{q^{ij}p_i p_j + m^2}
\lbl{2.15a}
\ee
\bear
  \delta N &:& {\cal H}^{\rm ADM} = \frac{1}{N}(p_0 - N^i p_i) \delta^3 (\bx - \bX)
  = \sqrt{q^{ij}p_i p_j + m^2} \delta^3 (\bx - \bX)  \lbl{2.13}\\
  \delta N^i &:& \frac{\alpha}{{\dot X}^0} p_i \frac{1}{N^2}(p_0 - N^j p_j) \delta^3 (\bx - \bX)
  = \frac{\alpha}{{\dot X}^0} p_i p^0 \delta^3 (\bx - \bX) = p_i \delta^3 (\bx - \bX) \lbl{2.14}
\ear
Here we simplified the notation so that now ${\dot X}^0 = {\dot X}^0|_{\tau_c}$.

The canonical momenta $p_\mu = \p L_{\rm m}^{(\tau)}/(\p {\dot X}^\mu)$, calculated from the action
(\ref{2.8}), whose matter part contains the parameter $\tau$, are the same as the canonical momenta
$p_i = \p L_{\rm m}^{(t)}/\p \sX^i$ and the quantity $p_0$ obtained from the action (\ref{2.10}) in which
$\tau$ was integrated out and the time parameter was $t$. This can be seen from the relations (\ref{2.11})--(\ref{2.15a}).

Equations (\ref{2.15}),(\ref{2.13})and (\ref{2.14})  imply the following
constrainsts\ci{PavsicKleinWheeler,PavsicKleinWheeler1}:
\bear
  &&\chi = \frac{1}{N^2}(p_0 - N^i p_i)^2-q^{ij}p_i p_j - m^2 = 0,  \lbl{2.19}\\
  &&\phi = {\cal H}^{\rm ADM} - \frac{1}{N} (p_0 - N^i p_i) \delta^3 (\bx - \bX) = 0 ,
  \lbl{2.17}\\
  &&\phi_i = {\cal H}_i^{\rm ADM} - p_i \delta^3 (\bx - \bX) = 0 .
  \lbl{2.18}
\ear

The Hamiltonian is a superposition of those constraints:
\be
  H = \int \dd^3 \bx \, \left ( \lambda \chi  \delta^3 (\bx - \bX) + N \phi + N^i \phi_i  \right )= 0 .
\lbl{2.20}
\ee
Using (\ref{2.19})--(\ref{2.18}), we obtain
$$
  H = \int \dd^3 \bx \, \left [ \lambda \left (\frac{1}{N^2}(p_0 - N^i p_i)^2-q^{ij}p_i p_j - m^2 
  \right ) \delta^3 (\bx - \bX) \right . \nonumber $$
\be  
    + N {\cal H}^{\rm ADM} + N^i {\cal H}_i^{\rm ADM}
  - p_0 \delta^3 (\bx - \bX) \biggr]  = 0 .
\lbl{2.21}
\ee
The terms with $N^i p_i$ have canceled out in the latter expression. The same Hamiltonian we also
obtain from the action (\ref{2.10}) according to the expression
\be
  H= \int \dd^3 \bx \, \left ( p_i \sX^i \delta^3 (\bx - \bX) + \pi^i \sq_{ij} - {\cal L} \right ) .
\lbl{2.20a}
\ee

From Eq.\,(\ref{2.21}), after using (\ref{2.19}) we obtain\ci{PavsicKleinWheeler1}
\be
   \int \dd^3 \bx \, \left ( N {\cal H}^{\rm ADM} + N^i {\cal H}_i^{\rm ADM} \right ) = p_0 .
\lbl{2.21a}
\ee
This corresponds to Eq.\,(\ref{2.7}), it is its ADM split analog.
   
The equations of motion obtained from the Hamiltonian (\ref{2.21}) are:
\bear
  &&{\mr p}_i = \lbrace p_i,H \rbrace = - \frac{\p H}{\p X^i}~,
  ~~~~\sX^i = \lbrace X^i,H \rbrace = \frac{\p H}{\p X^i} , \lbl{2.22}\\
   &&{\mathring{\pi}}^{ij} = \lbrace \pi^{ij},H \rbrace = - \frac{\delta H}{\delta q_{ij}}~,
  ~~~\sq_{ij} = \lbrace q_{ij},H \rbrace = \frac{\delta H}{\delta \pi^{ij}} , \lbl{2.23}
\ear
where the usual Poisson brackets relations have been used. The quantity $p_0$ is given by
Eq.\,(\ref{2.15a}), which comes from the constraint (\ref{2.18}).

Besides Eq.\,(\ref{2.22}), there is also the equation of motion for $p_0$, namely,
\be
  \frac{\p H}{\p p_0} = - \frac{\p L}{\p p_0}= 0 .
\lbl{2.25a}
\ee
Namely, the same equations (\ref{2.22}),(\ref{2.23}) also follow directly from the phase
space action (\ref{2.10}) according to the Euler-Lagrange equations
\bear
  &&\frac{\dd}{\dd t}\frac{\p L}{\p \sX^i} - \frac{\p L}{\p X^i} = 0~,
  ~~~\frac{\dd}{\dd t}\frac{\p L}{\p \mr{p}_i} - \frac{\p L}{\p p_i} = 0~,
  ~~~ - \frac{\p L}{\p p_0} = 0 , \lbl{2.25b}\\
  &&\frac{\dd}{\dd t}\frac{\delta L}{\delta \sq_{ij}} - \frac{\delta L}{\delta q_{ij}} = 0~,
  	~~~\frac{\dd}{\dd t}\frac{\delta L}{\delta \mr{\pi}^{ij}} - \frac{\delta L}{\delta \pi^{ij}} = 0 .
\lbl{2.25c}
\ear
Equations (\ref{2.22}) together with (\ref{2.25a}) are equivalent to the geodesic equation.
  
The same constraints (\ref{2.19})--(\ref{2.18}) also follow directly from the
action (\ref{2.8}) which contains the ``time'' parameter $\tau$ and the velocities ${\dot X}^\mu$.
Then alls quantities $p_\mu = (p_0,p_i)$ have the role of canonical momenta derivable  according
to $p_\mu = \frac{\p L}{\p {\dot X}^\mu}$ from such $\tau$-depended Lagrangian.
The Hamiltonian defined in terms of a superposition of those contraints is again given by Eq.\,(\ref{2.20}) in which the parameter $\lambda$ is replaced by another parameter, namely $\alpha$.
The equations of motion for $X^\mu$, $p_\mu$ are
\be
  {\dot p_\mu} = \lbrace p_\mu,H \rbrace~,~~~~{\dot X}^\mu =  \lbrace X^\mu,H \rbrace .
\lbl{2.26}
\ee
Explicitly this gives
\bear
  &&{\dot p}_\mu = - \frac{\p H}{\p X^\mu}
   = - \frac{\alpha}{2} \p_\mu g^{\alpha \beta} p_\alpha p_\beta 
   =  \frac{\alpha}{2} \p_\mu g_{\alpha \beta}\, p^\alpha p^\beta , \lbl{2.27}\\
  &&{\dot X}^\mu = \frac{\p H}{\p p_\mu} = \alpha p^\mu .  \lbl{2.28}
\ear
From the latter equations, after using $\alpha=\sqrt{{\dot X}^\mu {\dot X}_\mu}/m$, we
obtain
\be
  \frac{1}{\sqrt{{\dot X}^2}} \frac{\dd}{\dd \tau} \left ( \frac{{\dot X}_\mu}{\sqrt{{\dot X}^2}} \right )
  - \frac{1}{2} \p_\mu g_{\alpha \beta} \frac{{\dot X}^\alpha {\dot X}^\beta}{{\dot X}^2} = 0 ,
\lbl{2.29}
\ee
or equivalently,
\be
\frac{1}{\sqrt{{\dot X}^2}} \frac{\dd}{\dd \tau} \left ( \frac{{\dot X}^\mu}{\sqrt{{\dot X}^2}} \right )
  + \Gamma_{\alpha \beta}^{\,\mu} \frac{{\dot X}^\alpha {\dot X}^\beta}{{\dot X}^2} = 0 ,
\lbl{2.30}
\ee
which is the equation of geodesic.

We see that regardles of whether we start (i) from the original phase space action (\ref{2.8}), or,
(ii) from the action (\ref{2.10}), we obtain the same Hamiltonian (\ref{2.21}). In both cases
the Hamiltonian is a superposition of the constraints (\ref{2.17})--{\ref{2.19}), obtained by varying
the action with respect to the Lagrange multipliers $\alpha$, $N$ and $N^i$. In the second case, the
Hamiltonian can also be obtained by using the expression (\ref{2.20a}).

In general, the total variation of an action $I = \int \dd^4 x\, {\cal L}(\phi^a,\p_\mu \phi^a)$
that includes a change $\delta x^\mu$ of the boundary, is (see, e.g.\ci{BarutBook})
\be
  {\bar \delta} I = \int_R \dd^4 x \, \delta {\cal L} + \int_{R-R'} \dd^4 x \, {\cal L} =
  	 \int_R \dd^4 x \,\delta {\cal L} + \int_B \dd \Sigma_\mu \, {\cal L} \delta x^\mu .
\lbl{2.31}
\ee
Assuming that the equations of motion are satisfied, we have
\be
  {\bar \delta}I = \int_B \dd \Sigma_\mu \left ( \frac{\p {\cal L}}{\p \p \phi^a} \delta \phi^a
  + {\cal L} \delta x^\mu \right ) = \int \dd^4 x \,\p_\mu 
  \left ( \frac{\p {\cal L}}{\p \p \phi^a} \delta \phi^a + {\cal L} \delta x^\mu \right ) .
\lbl{2.32}
\ee
Here $\delta \phi^a = \phi'^a (x)-\phi^a (x)$. Introducing the total variation 
${\bar \delta} \phi^a =$  $\phi^a (x')-\phi^a (x)=$  $\delta \phi^a + \p_\mu \phi^a \delta x^\mu$, we
obtain
\be
 {\bar \delta}I = \int_B \dd \Sigma_\mu \left [ \frac{\p {\cal L}}{\p \p_\mu \phi^a} {\bar \delta}\phi^a
+ \left ( {\cal L} {\delta_\nu}^\mu - \frac{\p {\cal L}}{\p \p_\mu \phi^a} \p_\nu \phi^a  \right )
\delta x^\nu \right ] .
\lbl{2.33}
\ee

Let us consider the action (\ref{2.10}), identify $\phi^a = (X^i,q_{ij})$, and take coordinates in which
the surface element is $\dd \Sigma_\mu = (\dd \Sigma_0,0,0,0)$, $\dd \Sigma_0 = \dd^3 \bx$.
Then Eq.\,(\ref{2.33}) gives
\be
  {\bar \delta} I = \int_{t_1}^{t_3} \dd^3 \bx \left [ \frac{\p {\cal L}}{\p \sX^i} {\bar \delta} X^i
  + \frac{\p {\cal L}}{\p \sq_{ij}} {\bar \delta} q_{ij} +
  \left ( {\cal L} -  \frac{\p {\cal L}}{\p \sX^i} \sX^i - \frac{\p {\cal L}}{\p \sq_{ij}}
  \sq_{ij} \right ) \delta t \right ] .
\lbl{2.34}
\ee
The quantities ${\p {\cal L}}/{\p \sX^i} = p_i$ and   ${\p {\cal L}}/{\p \sq_{ij}} = \pi^{ij}$
are, respectively, the generators of the infinitesimal translations $X^i \rightarrow X^i +{\bar \delta} X^i$
and $q_{ij} \rightarrow q_{ij} + {\bar \delta} q_{ij}$. The expression in front of $\delta t$ is
just the negative of the Hamiltonian (\ref{2.20a}).

If we consider the original phase space action (\ref{2.8}) and perform the change $\tau \rightarrow
\tau + \delta \tau$, then we obtain
\be
  {\bar \delta}' I = \int \dd \tau \, \frac{\dd}{\dd \tau} \left ( \frac{\p L}{\p {\dot X}^\mu} \delta X^\mu
  + L \delta \tau \right ) = \int \dd \tau \, \left ( p_\mu {\bar \delta} X^\mu + 
  (L-p_\mu {\dot X}^\mu ) \delta \tau \right ) .
\lbl{2.35}
\ee
Here $p_\mu$ are the generators of the infinitesimal transformations 
$X^\mu \rightarrow X^\mu + {\bar \delta} X^\mu$, where ${\bar \delta} X^\mu = X'^\mu (\tau') - X^\mu (\tau) =\delta X^\mu + {\dot X}^\mu \delta \tau$. The quantity in front of $\delta \tau$ is the
generator of infinitesimal transformations $\tau \rightarrow \tau + \delta \tau$; it is equal to
$\frac{\alpha}{2} (g^{\mu \nu} p_\mu p_\nu - m^2 )$. This is the Hamiltonian for the relativistic
particle and it gives the correct equations of motion (namely, that of a geodesic).

We see that by considering the total variation of the action (\ref{2.8}) that includes a change of
$\tau$, we find not only that $p_i$ are the generators of the infinitesimal ``translations''
of $X^\mu$, $i=1,2,3$, but also that $p_0$ is the generator of the translations of $X^0$.
As $p_i$ do not vanish, also $p_0$ does not vanish; $p_\mu = (p_0,p_i)$ are the canonical momenta,
conjugated to the particle's position variables $X^\mu (\tau) = (X^0 (\tau),X^i (\tau))$. Those
variables are distinct objects than the spacetime coordinates $x^\mu = (x^o,x^i)$, $x^0 \equiv t$.
Because the particle is embedded in spacetime, in the action (\ref{2.8}) there occurs 
$\delta^4 (x-X(\tau))$, which says precisely that, namely that the particles is described by
a worldline $x^\mu = X^\mu (\tau)$.

Let us now follow the approach by Rovelli\ci{Rovelli1,Rovelli2} and see what do we obtain if instead of the
phase space action (\ref{2.8}) we use the action (\ref{2.1}), express the metric according
to Eq.\,(\ref{2.9a}), and fix the parameter so that $\tau= x^0 \equiv t$. The action (\ref{2.1})
then reads\footnote{In the approach considered by Rovelli, also a term due to a clock variable on
the particle's world line was included.}
\be
  I = m \int \dd t\, \sqrt{N^2 - ({\mr \bX} + \bN)^2} + 
  \int \dd t\, \dd^3 \bx \left ( \pi^{ij} \sq_{ij} - N {\cal H}^{\rm ADM} - N^i {\cal H}_i^{\rm ADM}
  		\right ) ,
\lbl{2.36}
\ee
where $({\mr \bX} + \bN)^2 = q_{ij} ({\mr \bX}^i + N^i) ({\mr \bX}^j + N^j)$.
We will also write $\bp^2 = q_{ij} p^i p^j$. The particle momentum is
\be
  \bp = - \frac{m ({\mr \bX} + \bN) }{ \sqrt{N^2 - ({\mr \bX} + \bN)^2}} ,
\lbl{2.37}
\ee
from which we have
\be
   N^2 - ({\mr \bX} + \bN)^2 = \frac{N^2 m^2}{m^2 + \bp^2}~,~~~{\rm and}~~~
   {\mr \bX} + \bN = \frac{\bp N}{\sqrt{m^2+\bp^2}} .
\lbl{3.37a}
\ee

The Hamiltonian is given by
\bear
  &&H= \bp \sX - L_m + \int  \left ( \pi^{ij} \sq_{ij} - {\cal L}_G \right ) \,\dd^3 \bx , \nonumber \\
  && \hs{5mm} = - \bN \bp - N \sqrt{m^2 + \bp^2}
   + \int \dd^3 \bx \, \left (  N {\cal H}^{\rm ADM} + N^i {\cal H}_i^{\rm ADM} \right ) .
\lbl{2.38}
\ear

If we vary the action (\ref{2.36}) with respect to $N$ and $N^i$, we obtain the following
constraints\ci{Rovelli1,Rovelli2}:
\bear
   &&{\cal H}^{\rm ADM} = \sqrt{m^2+\bp^2} \delta^3 (\bx - \bX ) , \lbl{2.39} \\
   &&{\cal H}_i^{\rm ADM} = p_i \delta^3 (\bx - \bX ) . \lbl{2.39}
\ear
The Hamiltonian (\ref{2.38}) is a superopsition of those constraints with the coefficients
$N$, $N^i$, and is therefore equal to zero (in the weak sense, i.e., on the constraint surface
in the phase space).

But if we consider the form (\ref{2.1}) of the action, before fixing $\tau$, then we obtain
the momentum $p_0 = {\p L}/{\p {\dot X}^0}$, besides the momenta $p_i = {\p L}/{\p {\dot X}^i}$.
Together all those momenta $p_\mu = (p_0,p_i)$ are constrained according to
\be
  g^{\mu \nu} p_\mu p_\nu - m^2 = \frac{1}{n^2} (p_0 - N^i p_i)^2 - q^{ij} p_i p_j - m^2 = 0 .
\lbl{2.41}
\ee
Solving the latter equation for $p_0$, we obtain Eq.\,(\ref{2.15a}), i.e.,
$p_0 = N^i p_i + N \sqrt{q^{ij} p_i p_j + m^2}$. Using the latter relation in Eq.\,(\ref{2.38}),
we obtain
\be
  H_G \equiv \int \dd^3 \bx \, \left (  N {\cal H}^{\rm ADM} + N^i {\cal H}_i^{\rm ADM} \right )
  = p_0 .
\lbl{2.42}
\ee
This means that the gravitational Hamiltonian is equal to the particle momentum $p_0$ which,
as we have seen before, is the generator of the transformation
$X^0 \rightarrow X^0 + {\bar \delta} X^0$.
Altogether, $p_\mu = (p_0,p_i)$ generate the transformations
$X^\mu \rightarrow X^\mu + {\bar \delta} X^\mu$, i.e., they shift the particle's position in spacetime.
Recall that Eq.\,(\ref{2.42}) is consistent with Eq.\,(\ref{2.7}) that we obtained directly
from Einstein's equations.
  
A different role has the $H$ of Eq.\,(\ref{2.38}). It is the generator of the transformation
$t \rightarrow t+ \delta t$ which in the passive interpretation is just a change of a coordinate,
a reparametrization. And because the action is invariant under reparametrizations of $x^\mu$, the
corresponding generators, defined in Eq.\,(\ref{2.33}), vanish, and so does the $H$
in Eq.\,(\ref{2.38}). 

According to the well known Einstein's hole argument, spacetime points cannot be identified.
A way to identify them is to fill spacetime with a reference fluid. If instead of a fluid we
have particles, then spacetime points are identified on the worldlines of those particles.
In the simplified model of a single particle, spacetime points are identified along the
worldline of that particle. From Eq.\,(\ref{2.35}) we
read that ${\bar \delta}X^\mu = ({\bar \delta}X^0,{\bar \delta}X^i)$ is a different sort
of transformation than $\delta x^\mu = (\delta x^0,\delta x^i)$, $\delta x^0 \equiv \delta t$.

An alternative approach was considered by Struyve\ci{Struyve1}. He started from the action
(\ref{2.1}) and cast it into such form that both terms, the particle's and the gravitational,
had the same ``time'' parameter $\tau$. For that purpose the integral
$\int \dd t\, \dd^3 \bx \, \sqrt{-g}\, R$ was transformed into
$\int \dd \tau \,{\dot X}^0 \, \dd^3 \bx\,\sqrt{-{\tl g}} \,{\tl R}$, where ${\tl R}$ and ${\tl g}$ were properly adjusted $R$ and $g$ that took into account the relation $x^0 = X^0 (\tau)$. Because of the occurrence of
${\dot X}^0$ in the gravity part of the action, the null-component of the canonical momentum was not
$p_0 = m {\dot X}_0/\sqrt{{\dot X}^2}$, but $p_0 = m {\dot X}_0/\sqrt{{\dot X}^2}+$  $\int \dd^3 \bx\,\sqrt{-{\tl g}}\, {\tl R}$.
Because of the extra term in $p_0$, the constraint no longer had the simple form (\ref{2.41}).
Consequently, instead of Eq.\,(\ref{2.42}) in which $p_0 \neq 0$, a different equation
was obtained, namely, $H_G =p_0$,
where $p_0$ turned out to be zero (i.e., it vanished weakly).
It was then concluded that, because of  the constraint $p_0 \approx 0$, a particle cannot
give rise to time in the quantum version of the theory and that the notorious problem of time still
existed. Struyve also observed that there exists a canonical transformation that relates the
approach based on the action (\ref{2.1}) to the approach based on his modified action, and that upon
quantization, those two approaches are related by the corresponding unitary transformations.

\section{Quantization}

We have arrived, following different paths, to the equation (\ref{2.42}), i.e.,
\be
   H= H_G - p_o = 0,
\lbl{3.1}
\ee
where $p_0$ is related to $p_i$ according to Eq.\,(\ref{2.15a}), whcih comes from the mass shell
constraint (\ref{2.15a}), and where 
$H_G = \int \dd^3 \bx \, \left (  N {\cal H}^{\rm ADM} + N^i {\cal H}_i^{\rm ADM} \right )$ contains the canonical momenta conjugated to the 3-metric $q_{ij}$.

Upon quantization, the particle coordinates $X^\mu$ and momenta $p_\mu$ become operators satisfying
\be
  [{\hat X}^\mu,{\hat X}^\nu] = 0~,~~~[{\hat p}_\mu,{\hat p}_\nu]=0~,~~~
  [{\hat X}^\mu,{\hat p}_\nu]=i {\delta^\mu}_\nu .
\lbl{3.2}
\ee
Similarly, the gravity variables become operators satisfying
\be
[{\hat q}_{ij},{\hat q}_{mn}] = 0~,~~~[{\hat \pi}_{ij},{\hat\pi}_{mn}]=0~,~~~
[{\hat q}_{ij},{\hat \pi}^{mn}]=i {\delta_{ij}}^{mn} .
\lbl{3.3}
\ee
In the Schr\"odinger representation in which $X^\mu$ and $q_{ij}$ are diagonal, Eqs.\,(\ref{3.2})
and (\ref{3.3}) are satisfied by ${\hat X}^\mu = X^\mu$, ${\hat q}_{ij} = q_{ij}$, ${\hat p}_\mu
= - i \p/\p X^{\mu}$ and $\pi^{ij} = -i \delta/\delta q_{ij}$.

The constraints (\ref{2.19})--(\ref{2.18}) become operator equations acting on the state vector,
which can be represented as a function of $X^\mu$ and a functional of $q_{ij} (\bx)$, namely
$\Psi [X^\mu,q_{ij} (\bx)]$.

In order to quantize Eq.\,(\ref{3.1}), we take the gauge $N=1$, $N^i =0$, and so we obtain
\be
  \int \dd^3 \bx \left ( \frac{-1}{\kappa} G_{ijk \ell} \pi^{ij} \pi^{k \ell} + \kappa
  \sqrt{q} R^{(3)} \right ) = p_0 ,
\lbl{3.3a}
\ee
where $\kappa = 1/(16 \pi G)$ and $G_{ijk \ell} \pi^{ij} \pi^{k \ell}$  Wheeler-DeWitt metric.
The latter equation can be written in the following compact form:
\be
  \frac{1}{\kappa} G^{a(\bx)b(\bx')} \pi_{a(\bx)} \pi_{b(\bx')} + V[q^{a(\bx)}] = - p_0 ,
\lbl{3.3b}
\ee
where $\pi_{a(\bx)} \equiv \pi^{ij} (\bx)$, $ G^{a(\bx)b(\bx')} = G_{ijk \ell} (\bx) \delta^3 (\bx-\bx')$,
$q^{a(\bx)} \equiv q_{ij} (\bx)$, and $V[q^{a(\bx)}] = - \kappa \sqrt{q} R^{(3)}$.
Upon quantization, $\pi_{a(\bx)}$ become the operators ${\hat \pi}_{a(\bx)} = -i \delta / \delta q^{a(\bx)}
\equiv - i \p_{a(\bx)}$, and $p_0$ becomes ${\hat p}_0 = -i \p/\p T$, where $T \equiv X^0$.

The notorious ordering ambiguity can be avoided\ci{PavsicOrder} if we proceed as follows. First, let me
illustrate the procedure  for the constraint (\ref{2.19}). It can be written in the following form:
\be
  \gam^\mu p_\mu \gam^\nu p_\nu - m^2 = 0.
\lbl{3.4}
\ee
Here $\gam^\mu$ are the generators of the Clifford algebra, satisfying
\be
   \gam^\mu \cdot \gam^\nu \equiv \frac{1}{2} (\gam^\mu \gam^\nu + \gam^\nu \gam^\mu ) = g^{\mu \nu} .
\lbl{3.5}
\ee
They have the role of the basis vectors\ci{Hesteness1,Hesteness2} (see also\ci{PavsicBook,PavsicKaluza,PavsicKaluzaLong}) in a curved spacetime $M$. The quantum version of Eq.\,(\ref{3.4})
is
\be
   \left (\gam^\mu {\hat p}_\mu \gam^\nu {\hat p}_\nu - m^2 \right ) \Psi = 0,
\lbl{3.6}
\ee
where $\Psi$ is a scalar wave function.
Using ${\hat p}_\mu  = -i \p_\mu$ and\footnote{Here the symbol $\p_\mu$ denotes a generic derivative that can	act on any Clifford algebra-valued field. For instance,
if acting on a scalar field it behaves as the partial derivative and if acting on basis vectors $\gam^\nu$
it determines the connection. More details and a justification why the same symbol $\p_\mu$ can be used in both
cases is provided in Ref.\ci{PavsicKaluzaLong}.}
	 $\p_\mu \gam^\nu =\Gam_{\mu \rho}^{~\nu} \gam^\rho$,
where $\Gam_{\mu \rho}^{~\nu}$ is the connection in $M$, equation (\ref{3.6}) becomes
\be
  \left (- {\Gam_{\mu \rho}}^{\,\nu} \gam^\mu \gam^\rho \p_\nu - \gam^\mu \gam^\nu \p_\mu \p_\nu - m^2 \right )
  	\Psi = 0 ,
\lbl{3.7}  
\ee
i.e.,
\be
\left (  \gam^\mu \gam^\nu \p_\mu \p_\nu + \Gam_{\mu \rho}^{\,\,\nu} g^{\mu \rho} \p_\nu + m^2 \right )
	\Psi = (\DD_\mu \DD^\mu + m^2 ) \Psi= 0 .
\lbl{3.8}
\ee
Here $\DD_\mu$ is the covariant derivative of the tensor calculus; acting on a vector components, it gives
$\DD a^\nu = \p_\mu a^\nu + {\Gam_{\mu \rho}}^{\,\nu} a^\rho$.

There is no ordering ambiguity in Eq.\,(\ref{3.6}), because $\gam^\mu {\hat \p}_\mu \gam^\nu {\hat \p}_\nu$ is the
product of two vector momentum operators\ci{PavsicOrder} ${\hat p} = \gam^\mu {\hat p}_\mu$ and is
invariant under general coordinate transformations. Using a different product, for instance,
${\hat p}_\mu \gam^\mu \gam^\nu {\hat p}_\nu$
would make no sense, because such an object is not invariant and not a product of two vector operators.
In Ref.\ci{PavsicStumbling} it was shown what happens if $\Psi$ in Eq.\,(\ref{3.6}) is a spinor field, expanded
in term of the spinor basis $\xi_\alpha$, according to $\Psi = \psi^\alpha \xi_\alpha$. Then, instead of
(\ref{3.8}) one obtains
\be
  ({\cal D}^\mu {\cal D}_\mu +m^2) \psi^\delta + \frac{1}{2} {[\gam^\mu,\gam^\nu]^\delta}_\alpha {R_{\mu \nu}^{~~\alpha}}_ {\beta} \psi^\beta,
\lbl{3.8a}
\ee
where ${\cal D}_\mu$ contains also the spin connection, determined by the relation $\p_\mu \xi_\alpha =
{{\Gam_\mu}^{\,\beta}}_\alpha \xi_\beta$. Analogously we can find the corresponding equation for a vector field.

In a similar way also the ordering ambiguity in Eq.\,(\ref{3.3b}) can be avoided by introducing the superspace
analog of $\gam^\mu$ and rewrite Eq.\,(\ref{3.3b}) in the form
\be
  \frac{1}{\kappa} G^{a(\bx)} \pi_{a(\bx)} G^{b(\bx')} \pi_{b(\bx')} + V[q] = - p_0 ,
\lbl{3.9}
\ee
where $ G^{a(\bx)}$ are the generators of the Clifford algebra in the superspace ${\cal S}$, satisfying
\be
   G^{a(\bx)} \cdot  G^{b(\bx')}
    \equiv\frac{1}{2} \left ( G^{a(\bx)} G^{b(\bx')} +  G^{b(\bx')} G^{a(\bx)} \right ) = G^{a(\bx)b(\bx')} .
\lbl{3.10a}
\ee

The quantum version of Eq.\,(\ref{3.9}) is
\be
 \left (  \frac{1}{\kappa} G^{a(\bx)} {\hat \pi}_{a(\bx)} G^{b(\bx')} {\hat \pi}_{b(\bx')} + V[q] \right ) \Psi
   		  = i \frac{\p \Psi}{\p T},
\lbl{3.10b}
\ee
where $\Psi = \Psi [T,X^i,q^{a(\bx)}]$. In the latter equation ${\hat \pi}_{a(\bx)} = - i \p_{a(\bx)}$ acts
on $G^{b(\bx')}$. Analogously as in the finite dimensional case, the derivative
$\p_{a(\bx)}$ acting on $G^{b(\bx')}$ gives the connection according to
\be
  \p_{a({\bx})} G^{b(\bx')} = \Gamma_{a(\bx) c(\bx'')}^{~b(\bx')} G^{c(\bx')}.
\lbl{3.10c}
\ee
Equation (\ref{3.10b}) then becomes
\be
  \left ( \frac{1}{\kappa} G^{a(\bx) b(\bx)} \p_{a(\bx)} \p_{b(\bx')}
   + \Gamma_{a(\bx) c(\bx'')}^{~b(\bx')} G^{a(\bx) c(\bx'')} \p_{b(\bx')} + V[q] \right ) \Psi
 =  \left ( \DD_{a(\bx)} \DD^{a(\bx)} + V [q] \right ) = i \frac{\p \Psi}{\p T} .
\lbl{3.11}
\ee
The connection is given by
\be
    \Gamma_{a(\bx) b(\bx')}^{~c(\bx'')} = \frac{1}{2} G^{c(\bx'') d(\bx''')}
    \left ( G_{a(\bx) d(\bx'''),b(\bx')} + G_{d(\bx''') b(\bx''),a(\bx)}
     -  G_{a(\bx)b (\bx'),d(\bx''')} \right ) ,
\lbl{3.12}
\ee
where the comma denotes the functional derivative. Using the established techniques for the
superspace calculations (see, e.g.,\ci{Superspace}), the connection (\ref{3.12}) can be calculated  for the
Wheeler-DeWitt metric by using 
\be
 \p_{c(\bx'')} G^{a(\bx) b(\bx')} = \frac{\delta}{\delta q_{mn}} G_{ijk\ell} \delta (\bx-\bx')~,
 	{\rm and} ~~~\frac{\delta}{\delta q_{mn}} g_{ij} (\bx') = \delta_{(i}^{(m} \delta_{j)}^{m)} .
\lbl{3.13}
\ee
Because the ordering issues regarding the Wheeler-DeWitt equation are not the main topics of this
paper, they will be discussed in more detail elsewhere.

\section{Conclusion}

We have considered the relativistic particle coupled to gravity and analysed the constraints satisfied
by such system. The constraints follow directly from the action (\ref{2.1}) by varying it with respect
to the non dynamical components of the metric $g^{\mu}$, namely, $g^{0\mu}$, which gives
the $(0 \mu)$-components of Einstein's equations: $\phi_\mu = \frac{1}{8 \pi G} G_{o \mu} + T_{0 \mu} = 0$.
The Hamiltonian is a linear superposition of those constraints,
$H = \int \alpha^\mu \phi_\mu \sqrt{-g} \dd^3 \bx = H_g + H_m = 0$, where
$H_m = \int T_{0 \mu} g^{0 \mu} \sqrt{-g} \dd^3 \bx = p_0$ is the particle momentum. If we perform
the ADM split of the action (\ref{2.1}) or its phase space form (\ref{2.8}), then the non dynamical
variables are the lapse and shift functions, $N$ and $N^i$. The constraints $\phi$, $\phi_i$ come
from varying the action with respect to $N$ and $N^i$. The time variable in the matter and the gravity
part of the action are the same, namely $t\equiv x^0$. In addition, the matter part $I_m$ contains
the worldline parameter $\tau$ and the term $\delta^4 (x-X(\tau))$ which tells that the worldline
is embedded in spacetime and thus satisfies the parametric equation $x^\mu = X^\mu (\tau)$. Because
$I_m$ is invariant under reparametrizations of $\tau$, the particle momenta $p_\mu = \p L/\p {\dot X}^\mu$
satisfy the mass shell constraint, $\chi = p_\mu p^\mu - m^2 = 0$. The Hamiltonian, which is a superposition
of the constrainst $\chi$, $\phi$ and $\phi_i$ gives the correct equations of motion for all the dynamical
variables by using the ordinary Poisson brackets. The matter part of the Hamiltonian is $p_0$, as
it should be, because---as we have seen--- it also comes directly from the Einstein equations.

Upon quantization, $p_0$ becomes the operator ${\hat p}_0 = i \p/\p T$, where $T \equiv X^0$ donotes
the time coordinate of the particle. Altogether, $X^\mu$, $\mu= 0,1,2,3$, denote position of the
particle in spacetime. The Hamilton constraint $H=H_g + H_m =0$, i.e., $H_g = -p_0$, becomes the
Schr\"odinger-like equation $H_g  \Psi = i \p \psi/\p T$, where $\Psi = \Psi[T,X^i,q_{ij}]$. The time
and hence spacetime in this approach does not disappear upon quantization.

Spacetime in the quantized theory disappears if no matter is present. 
According to Einstein's hole
argument, spacetime points cannot be identified. They can be identified in the presence of a reference fluid.
If there is no reference fluid and instead there are particles, then spacetime points can be identified
on the worldlines of the particles. In this paper we have considered a simplified model with only one particle
present, and found that its time coordinate $T$ remains in the quantized theory.
In the usual approaches in which matter is given by some fields, such as a scalar or spinor field, spacetime
points cannot be directly identified as in the case of particles and one faces the notorious problem of
time, namely, how those fields could give rise to time.
 There is a lot of discussion of how to resolve it, but so far no generally accepted resolution has been
found. In the model in which gravity is coupled to particles, the problem of time does not exist.

\end{document}